# Brace for Impact: A Review of Mitigation Decisions of Critical Infrastructure Operators During the 2024 Solar Maximum

C. M. LaNeve[1], E. J. Oughton[1], N. Rivera[1], L. Wilkerson[2], R. S. Weigel[2], D. Thomas[2] and C.T. Gaunt[3]

[1]Geography and Geoinformation Science Department, George Mason University, Fairfax, VA, USA
[2]Physics & Astronomy Department, George Mason University, Fairfax, VA, USA
[3]Dept. Of Electrical Engineering, University of Cape Town, Cape Town, ZA

Corresponding author: Caitlin LaNeve (claneve@gmu.edu)

**Key Points:**

- This study records perishable information from critical infrastructure operators, benchmarking future studies and procedures.

- Operators favor modeling and situational awareness over the expensive hardware upgrades frequently recommended in literature.

- Organizations prioritize mission continuity for revenue, often bypassing risk-averse advice such as safe mode or asset maneuvers.

## Abstract

The Gannon Storm in May 2024 was the largest space weather event experienced in 20 years, generating auroras latitudes as low as 35°. Such activity can pose significant operational challenges for critical infrastructure operators, particularly those managing electricity transmission networks, satellite constellations, and aviation systems. Substantial progress has been made in understanding space weather and the potential exposure of infrastructure assets to severe events. We have seen few evaluations of the types of current mitigation strategies in use to reduce our shared vulnerability to this activity, motivating this study. Firstly, we fill an important literature gap by undertaking a systematic review of the range of space weather mitigation strategies for these three critical infrastructure sectors. Secondly, we contacted 303 critical infrastructure operators (50 power, 227 satellite, 26 aviation) for participation in an anonymous online survey or interview receiving 55 unique responses (18% response rate). To capture narratives of mitigation actions taken and impacts experienced over the solar maximum, qualitative interviews were then conducted with 33 operators. The results identified 91 potential mitigation actions within the sectors and found that 149 mitigation actions were enacted due to space weather forecasts and experienced impacts. This is one of the first exhaustive studies of space weather mitigation activities, and moves beyond the traditional focus on purely impacts.

**Plain Language Summary**

This study focuses on how power, satellite, and aviation operators handled the massive space weather event, Gannon Storm, to see if real-world actions matched academic advice. While research often suggests large purchases of expensive hardware, or shutting down systems to avoid damage, most companies opted to continue operations to avoid losing money, relying on modeling and alert systems to manage risk. We identify a preparedness gap where larger companies have the ability to harden networks before operations, but smaller ones may choose to weather the storm due to high costs. Ultimately, this study highlights that instead of only purchasing new equipment, operators identified the need for mor accurate, localized, space weather warnings from the government to protect our critical infrastructure.

## 1 Introduction

Space weather arises from many different types of solar phenomena, including solar flares, Coronal Mass Ejections (CMEs), and solar proton events, presenting hazards to assets on orbit, as well as technology and infrastructure on Earth's surface (Xue *et al.*, 2024a). The Sun has a cycle of increased and decreased solar activity over approximately 11 years, or solar maximum and minimum periods (Cao, Feng and An, 2022). The solar maximum is caused by the Sun's magnetic poles flipping and leads a period of increased activity associated with frequent solar flares and sunspots (Interrante, 2024, Jones et al., 2020). The reversal in alignment has higher chances of disruption associated with space weather, including but not limited to impacts to power grids, satellite operations, and global positioning and navigation (Maletckii and Astafyeva, 2024; Clilverd et al., 2025a; Kashcheyev et al., 2025). This paper includes a review on how system operators across multiple industries react to solar events, policies that impact operations, and a review on operator decisions during the current solar maximum in the mid-2020s.

The geomagnetic storm of May 10th, 2024, named the "Gannon Storm," was the first time since the 2003 Halloween storms that the US National Oceanic and Atmospheric Administration (NOAA) Space Weather Prediction Center (SWPC) issued an Extreme G5 geomagnetic event warning, the highest rating on the scale (Evans et al., 2024). Extreme space weather events provide crucial data for modeling and prediction purposes, with many using data from the Carrington superstorm of 1859 and the 2003 Halloween storm for validation (Ngwira et al., 2014; Blake et al., 2021). Therefore, the importance of this event provides strong motivation to collect perishable operational decision data from critical infrastructure operators, especially in sectors of high infrastructural importance.

To boost societal resilience to space weather threats we need to be fully aware of the range of potential mitigation decisions available to critical infrastructure operators. We also need to collect, analyze and document information on space weather impacts that have taken place over the solar maximum. Subsequently, this study explores the following three research questions:

1. What is the current range of mitigating decisions available to critical infrastructure operators to protect against space weather, as identified in the current literature?
2. Which of these decisions were carried out over the current solar maximum (cycle 25), including during the major Gannon and October 2024 storms?
3. How do the mitigation decisions identified in literature compare to the decisions made by critical infrastructure operators during solar events?

## 2 Literature review

In this section we undertake a systematic review of the literature focusing on three major critical infrastructure sectors, specifically electricity transmission (power), satellite constellations, and aviation organizations.

## 2.1. Electricity Transmission (Power) Infrastructure

Space weather poses a significant risk to power grids and extremely high voltage (EHV) transformers due to the creation of geomagnetically induced currents (GICs), resulting from geomagnetic disturbances (GMDs) (Cordell et al., 2024; Smith et al., 2024; Vandegriff et al., 2024; Oughton et al., 2025). Charged particles emitted from the Sun during these CMEs can interact with Earth's magnetic field (Owens et al., 2021; Lanabere et al., 2023; Liu et al., 2024; Pulkkinen et al., 2025). The disturbances of the magnetosphere and ionosphere induce electric fields upon Earth's surface, leading to low-frequency GICs flowing into conductive networks such as electricity transmission infrastructure (Ingham et al., 2023; Ngwira et al., 2023; Pratscher et al., 2024; Thomas et al., 2024).

Power systems support the functionality of all other infrastructure systems, highlighting their importance (Hall et al., 2016, 2017). Generally, the GICs a power grid can be exposed to are highly dependent on the voltage, configuration, length and spatial direction of the network (Andreyev et al., 2023). Recently, research has shown that substations located in high ground resistivity areas or near long transmission lines experience heightened GIC impact due to stronger induced geoelectric fields (Kazerooni, Zhu and Overbye, 2017; Barnes, Mate and Bent, 2024; Oughton et al., 2025a). This emphasizes the importance of understanding the three-dimensional Earth conductivity structure in Earth's crust when trying to model GICs (Nakamura et al., 2018; Espinosa et al., 2023; Marshalko et al., 2023). In terms of network configuration, longer and higher voltage transmission lines tend to conduct higher GICs, with the potential to flow into substations and increase vulnerability to space weather (Kazerooni, Zhu and Overbye, 2017; Caraballo et al., 2023). Susceptibility can also vary by orientation, with transmission lines parallel to the Earth's magnetic field being up to several times more affected by GICs than those oriented perpendicular to it, leading to uneven impacts across the grid (Mayer and Stork, 2024).

As GICs flow through power transmission systems, critical components such as EHV transformers are especially exposed because they have low resistance to direct current (DC)-like currents, such as GICs, and can cause saturation of the transformer's core, leading to inefficient operation and overheating (Ahmadzadeh-Shooshtari and Rezaei-Zare, 2022a; Subritzky et al., 2024). Three main mechanisms of failure from GICs are identified. Firstly, prolonged saturation can damage transformer windings, leading to permanent failure from transformer heating (Albert et al., 2022a; Akbari and Rezaei-Zare, 2023; Akbari, Mostafaei and Rezaei-Zare, 2023). Secondly, inefficient power delivery, often referred to as excessive reactive power being drawn from the grid, can lead to voltage instability, potentially leading to a voltage collapse (Saleh et al., 2024).

Thirdly, harmful harmonics can be produced that can affect grid operations (Ahmadzadeh-Shooshtari and Rezaei-Zare, 2022b; Crack et al., 2024) and cause important components such as transformers to trip, removing an asset from normal operation (Dimmock et al., 2024). There is concern that any of these mechanisms can lead to a GIC-induced cascading failure with widespread power outages, as seen during the 1989 Quebec blackout (Boteler, 2019; Oughton et al., 2024).

Given this, a variety of mitigation decisions have been identified from the literature and visualized within Figure 1. Long-term mitigation strategies include studying operational responses to extreme voltage fluctuations and conducting vulnerability modeling to understand how GICs impact the system (U.S. Department of Energy (DOE), 2019; Clilverd et al., 2025b). Implementing cooling mechanisms to mitigate transformer heating and applying protective materials to smaller grid assets help reduce long-term degradation (Albert et al., 2022b). The development of transformer replacement plans and purchasing backups for vulnerable components ensure the availability of critical infrastructure during emergencies (Mac Manus et al., 2023). Additionally, recording baseline transformer currents and conducting benchmark GMD Vulnerability Assessments every 60 calendar months allows operators to track changes over time and refine predictive models (NERC, 2019; U.S. Department of Energy (DOE), 2019). Reviewing past GIC events to improve simulations, modifying systems to mitigate harmonic distortions, and ensuring the availability of spare transformers further strengthen resilience (Cordell et al., 2025). Common mitigations involve grid sequencing, grid islanding, and installing GIC blockers (Oughton *et al.*, 2025) These efforts, combined with black-start preparedness and relay current adjustments, enable utilities to withstand and recover from severe space weather disturbances (Alassi et al., 2021).

Short-term mitigation decisions are implemented when an extreme space weather event is imminent, and a space weather forecast has been issued for an incoming event (Smith et al., 2022). These actions are designed to maintain grid stability and prevent widespread failures (Etchells *et al.*, 2024a). Activating GIC blocking devices and increasing reactive power reserves help counteract the impacts of GMDs (Patel, 2024). Operators alert staff to prepare emergency procedures, cancel employee time off to ensure adequate workforce availability, and implement strategic line rearrangements to distribute loads efficiently (NERC, 2012). Bringing additional generation capacity online and taking vulnerable assets offline can help balance the system load (Mac Manus et al., 2023), while temporarily canceling network maintenance frees up labor for crisis response. Increasing situational awareness among staff, bringing assets under maintenance back online, and revising preparation plans in real time further enhance the system's ability to respond dynamically to the event (Jasiūnas, Lund and Mikkola, 2021).

Post-event, utilities must evaluate affected assets, investigate unusual observations, and undertake modeling activities to refine future response strategies (NERC, 2012). Key steps include assessing dissolved gas levels in exposed transformers, replacing damaged EHV assets, and removing compromised components from service for inspection (Song et al., 2022). Sharing experiences and data with other operators contributes to improved industry-wide resilience (Gonzalez-Esparza et al., 2024).

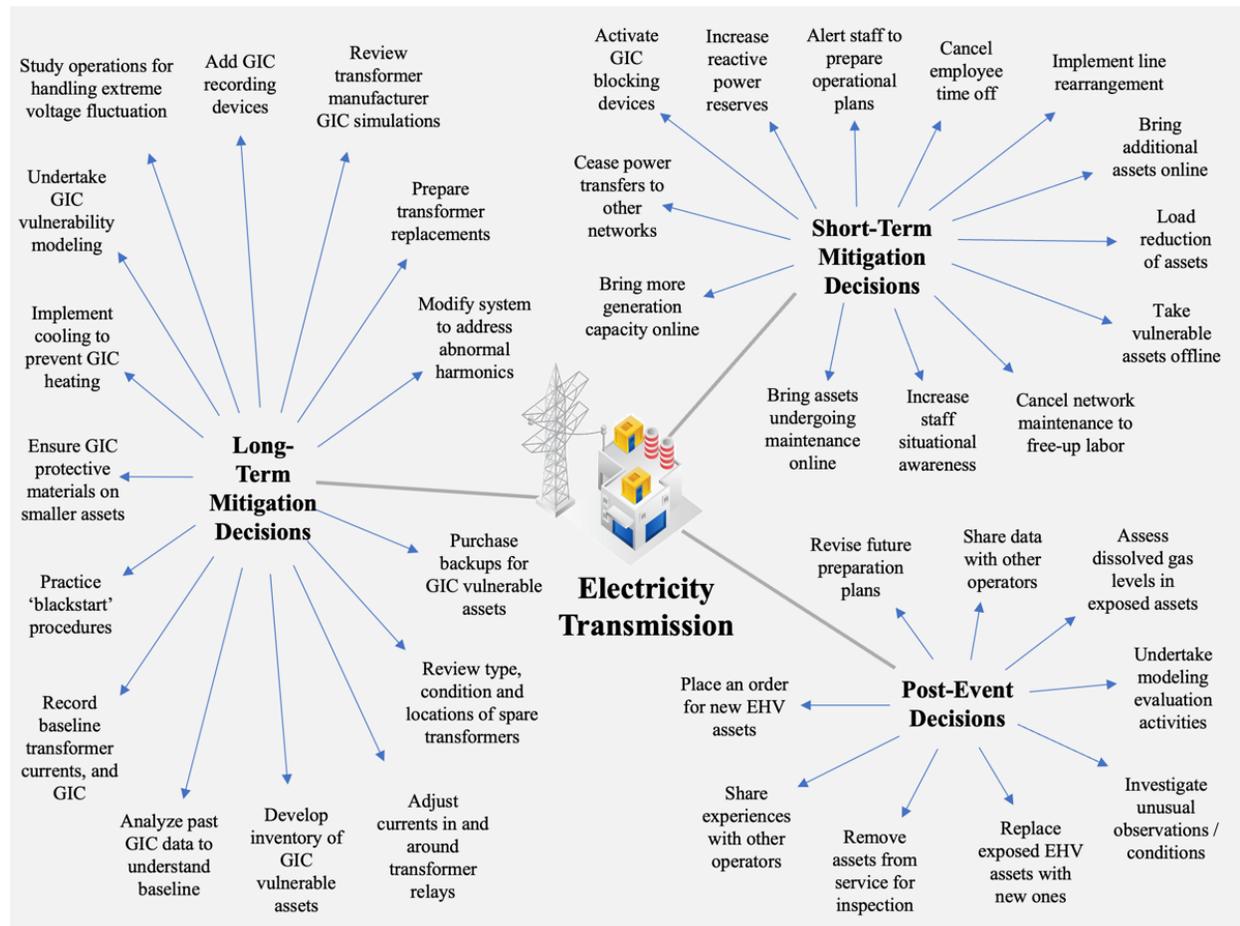

**Figure 1.** Summarizing pre- and post-event mitigation decisions for power grids

## 2.2. Satellite constellations

Solar activity has a significant impact on satellites in orbit, including potential disruption of communications, surface charging of electronics, and changes to orbital drag (Gabrielse et al., 2022; O'Brien et al., 2022; Aleshko et al., 2023; Yue et al., 2023). This solar radiation can also cause single event upsets (SEUs) across a variety of electronic components commonly used in constellations (Zhang et al., 2023).

Satellite operators are faced with a plethora of information, limited time for decision making, and limited knowledge of historical space weather relative to their other areas of expertise, causing the need for confident forecasts (Boyd et al., 2023; Feng et al., 2023). An operator will prioritize the safety of the satellite equipment, resulting in transferring the asset into safe mode (Qiao et al., 2024; Arif, 2025). Space weather events, such as geomagnetic storms, can affect the magnetic field of the Earth and the structure of the upper atmosphere by a process called Joule Heating. When CMEs interact with Earth's geospace environment they can deposit large quantities of energy into the ionosphere, such that kinetic energy is converted to thermal energy, and the total density of the ionosphere increases (Smith et al., 2023; Parker and Linares, 2024). The result of this is that Low Earth Orbit satellites experience more drag in their holding pattern by the increased weight of their occupying atmosphere (Fang et al., 2022a; Yin et al., 2024; Sanchez-Hurtado et al., 2025). Current research indicates that these problems become worse on an increase in total satellite launches (aided by companies like Starlink) and debris fields from fragmented spacecraft necessitate the implementation of effective "collision avoidance" (Parker and Linares, 2024; Parker et al., 2024; Kukreja, Oughton and Linares, 2025; Osoro et al., 2025).

The second main hazard indicated in the literature is radiation in the form of both ionized and non-ionized radiation. Total ionizing dose (TID) is a slow and compounding issue (Hands et al. 2018). Ionization within the structure of a satellite leads to the accumulation of positive charges, leading to critical failure in components (Yang et al., 2022, Inchin et al., 2025, Berger et al., 2023). Displacement damage dose (DDD) is the cumulative degradation of materials, mostly caused by solar protons and neutrons (Iwamoto and Sato, 2022). Damage to satellites would likely be manageable, regardless, researchers call upon more studies in this area (Gozutok and Kaymaz, 2023).

The third main hazard are SEUs, moments in which a satellite fails in single instances due to any hazard or stimuli (Etchells et al., 2024b). This hazard can occur with a radiated space environment and the pressure of high doses of high-energy trapped electrons in Earth's magnetic field being released on satellites during a space weather event, subsequently putting them at risk of severe life-long damage. This hazard manifests in satellites in two ways: TID and DDD (Hands et al., 2018; Xue et al., 2024b). However, because Space Weather can induce all three of these environmental effects on a satellite, it is easy to apply the same mitigation procedures taken by a satellite operator in response to these hazards to forecasting of a space weather event. Figure 2 visualizes the range of mitigation options identified for the satellite sector.

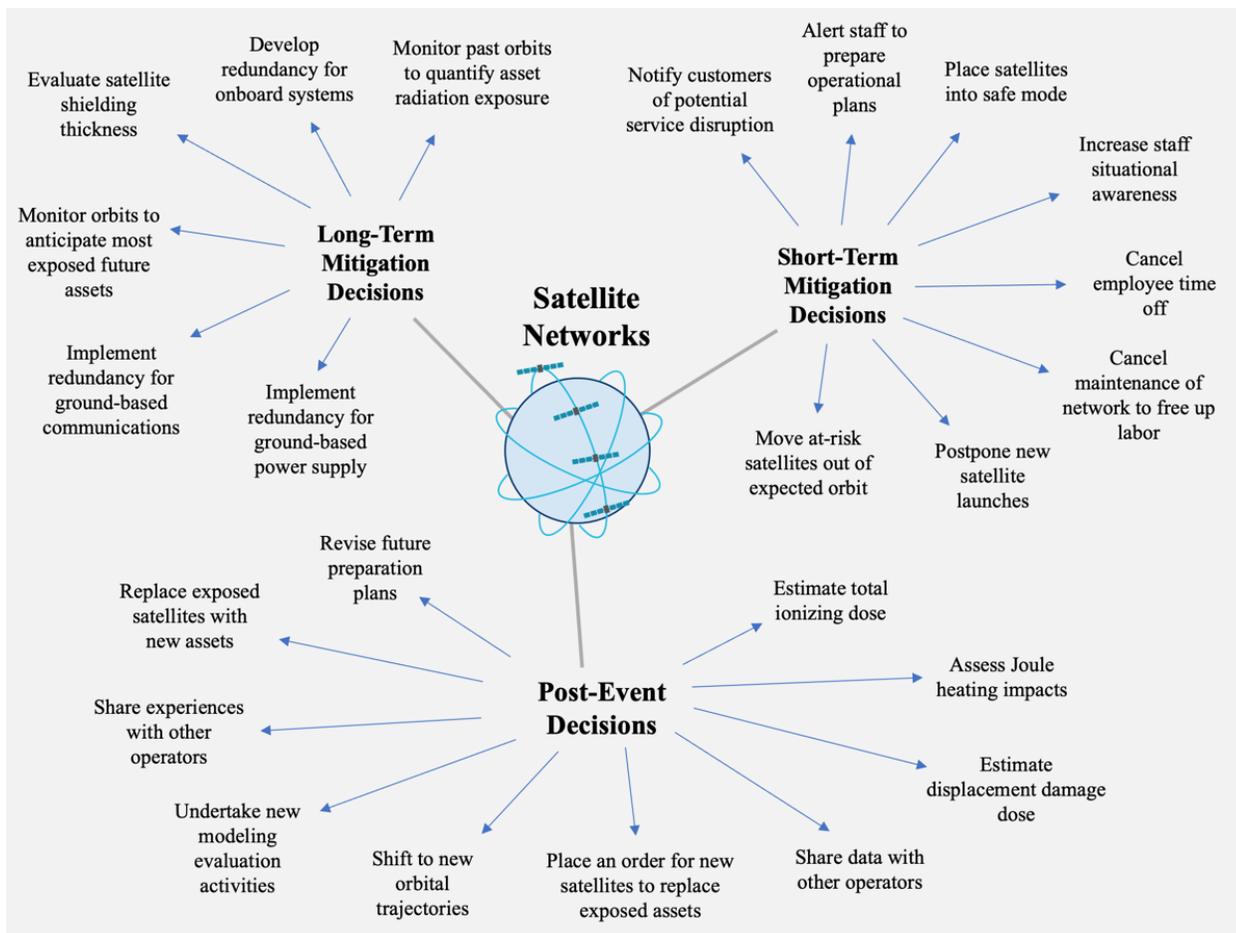

**Figure 2.** Summarizing pre- and post-event mitigation decisions for satellite constellations

Long-term mitigation decisions focus on proactive measures to reduce the risks posed by space weather over time. Importantly, evaluating satellite shielding thickness for adequate protection against radiation exposure (Shah et al., 2024). Monitoring past orbits helps quantify asset radiation exposure, allowing operators to assess risks and adjust designs for future missions (Parker and Linares, 2024). Monitoring future orbits enables operators to predict which assets will be most exposed and make necessary adjustments (Spence et al., 2022; Miteva, Samwel and Tkatchova, 2023). Implementing redundancy for onboard systems, ground-based communications, and power supply ensure that satellite operations can continue (Leahy, 2021; Praks et al., 2023; Zhang et al., 2023). Developing backup strategies, such as replacing exposed satellites with new assets and revising preparation plans based on past incidents, enhances long-term resilience against extreme space weather events (Qiao et al., 2024).

Short-term mitigation decisions focus on minimizing potential damage and maintaining service continuity. Operators alert staff to prepare operational plans and notify customers of possible service disruptions (Blumenfeld, 2024). At-risk satellites may be moved out of their expected orbits to reduce exposure, while placing satellites into safe mode protects sensitive electronics from excess radiation (Mahmood et al., 2025; Shambaugh, 2025). Increasing situational awareness among staff, canceling employee time off, and freeing up labor by postponing maintenance and new satellite launches ensure that adequate personnel are available to manage the crisis (Fang et al., 2022b). These rapid response measures help prevent irreversible damage and sustain critical functions during space weather events.

Post-event decisions involve assessing and mitigating the damage sustained from space weather disturbances. One crucial step is estimating the TID absorbed by satellite components, which helps determine their remaining lifespan and performance capacity (Leahy, 2021; Serief and Meguenni, 2025). Assessing Joule heating impacts and estimating DDD further contribute to understanding the extent of degradation (Parker 2024). Damaged satellites may be replaced with new assets, and orbital trajectories may be adjusted to optimize future satellite positioning (Praks et al., 2023). Sharing data and experiences with other operators enhances collective knowledge, leading to improved mitigation strategies (Arif, 2025).

### 2.3. Aviation

Hash radiation environments can cause harm to avionics and aviation personnel, especially during longer flights and at altitudes closer to the ionosphere (Knipp, 2017). Ionizing radiation is an identifiable hazard in this environment and can take two forms: galactic cosmic radiation and solar energetic particle events (Phoenix et al., 2024). The NOAA SWPC conducted a recent survey underscoring the importance of better standards, procedures, and forecasting in the aviation sector (Bain et al., 2023).

Based on a study conducted of 4 million flight data records, space weather events correlated in an 81.34% uptick of flight arrival delays and 21.45% of 30-minute delays (Wang et al., 2023). Avionic electronic systems are affected by solar energetic particles that originate from space weather events (Buzulukova and Tsurutani, 2022). As certain aspects of aviation like communication systems are vital to the safe operation and process of a flight from beginning to end, any deviation from these normal procedures warrants a delay. A second study found that the mean departure delay time increased by 20.68% (7.67 minutes) compared to times with no solar flare events (Xu et al., 2023).

The Partnership for Excellence in Civil Aviation Space Weather User Services (PECASUS) authored a document summarizing Space Weather advisories that are given to pilots and monitors space weather conditions regarding its effect on HF Communication and satellite navigation (PECASUS, 2020). If space weather events threaten these systems, the International Civil Aviation Organization (ICAO) sends advisories to flight crew (Hubert and Aubry, 2021). This PECASUS document gives further insight into current practices by which aviation professionals can be aware of possible hazards during their flight. Figure 3 visualizes the range of mitigation options identified for the aviation sector.

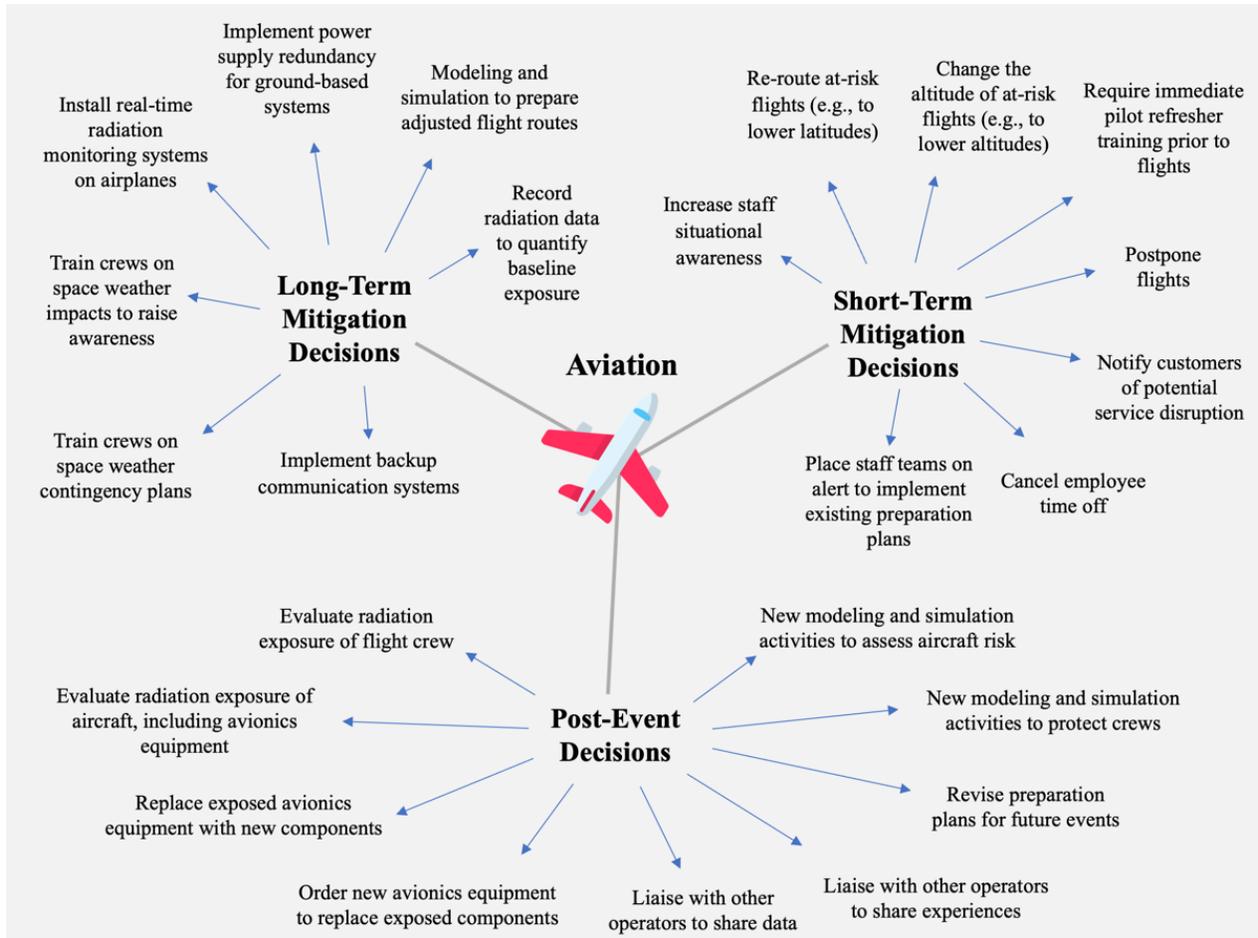

**Figure 3.** Summarizing pre- and post-event mitigation decisions for aviation

Solar radiation is an unavoidable potent consequence of space weather, thus the ability to mitigate the radiation itself is limited (Marov, 2020). For long-term mitigation, the American National Weather Service publishes radio blackout, solar radiation, and geomagnetic storm impact alerts that cover an 18 hour period of time (Sanders et al., 2017). Most mitigation strategies are focused on actions following a warning of a space weather event (Tezari et al., 2022). In terms of short-term mitigation decisions, it is first recommended that flights that can be delayed should be delayed, decreasing the time spent in an irradiated atmospheric environment (Xu et al., 2023), then consider flying at lower altitudes and latitudes, i.e.; changing their flight path, to better avoid elevated radiation exposure (Schmölter and Berdermann, 2024).

The ICAO has a network of space weather centers that share advisories and other information with one another, crucial in mitigation as information must be communicated to air traffic controllers around the world so that they can notify aircraft of new flight patterns, delays, and restrictions (Meier et al., 2020). With these forecasts and warning in hand, ICAO advises industry professionals to focus on areas that are known to be vulnerable to space weather events, including areas of high latitude and higher flight levels (Kauristie et al., 2021).

Chapter 6 of the FAA's Risk Management Handbook details the hazards pilots encounter in flight, and how they are trained to deal with them. Any type of hazard, whether damage to communication or navigation systems or otherwise, is to be dealt with through "Checklists, Standard Operating Procedures, and Best Practices" which a pilot is trained in by default (FAA, 2022). The checklists and standard operating procedures are what helped create several of our mitigation strategies noted in Figure 3, as they directly correlate with mitigation procedures in-flight.

Throughout their careers, crew members are recommended to track their exposure to radiation through published dosage calculators (Phoenix et al., 2024, FAA 2014). CARI-6, the calculator for radiation, derives the dose of galactic cosmic radiation through accounting for changes in altitude and geographic location during a flight as entered by the user. Past this, the FAA only mentions modeling and simulation post events (FAA, 2025).

## 3   Methodology

The literature review conducted in the previous sections identified the core mitigation techniques for reducing technology impacts during geomagnetic events from the perspective of critical infrastructure operators. To verify the results of the literature review and collect information on possible impacts over the solar maximum, we engage current operators in these sectors. Gathering this data from the primary source is imperative to understand how to improve mitigation techniques for all sectors. The survey is hosted on a website (www.chrono-storm.com), allowing operators in the power, satellite and aviation industries to anonymously complete the questions. The questions pertain to the protocols, observations, and decisions taken before, during, and after space weather events have taken place.

The survey will be separated into sectors: Electricity, Satellite, and Aviation users. Once a user selects the sector and agrees to participate, they will be shown a series of questions, which include Operator Background, Long-Term Pre-Event Preparation, Short-Term Pre-Event Preparation, Operational Decisions During an Event, Post-Event Decisions, Incidents, and Interview Opportunities. The survey itself has multiple options for each question along with free-text boxes to record additional information. The demographic information to be included in the survey will vary based on the sector, but will include basic information concerning the larger organization, not the participant, retaining participant privacy.

This study follows a population census approach, since the available population of operators with the necessary specific context is relatively limited (Willie, 2024). Thus, companies will be identified, contacted for participation, and tracked to avoid redundant responses. In all, 50 power, 26 aviation, and 227 satellite organizations were identified for participation in the study.

Participants will be given three options: conduct the online survey only (22), conduct the qualitative interview only (25), or conduct both the online survey and qualitative interview (8). The qualitative interviews allow for a greater depth of understanding from the on-the-ground perspective of sector professionals during May 2024 and over the solar maximum. No videos are to be stored from these interviews and summary transcripts will be scrubbed for identifying information, ensuring anonymity for the participants. The survey data is compiled and stripped of demographic information participants provided. Obtained data is organized according to the number of responses per answer.

## 4    Results

This section reports the results obtained from the survey and interview activities.

### 4.1. Power results

Figure 4 presents survey responses from power grid operators regarding mitigation strategies undertaken in anticipation of potential GMDs, as Figure 5 presents the interview responses. The horizontal bar chart categorizes responses based on the size of the organization, measured by the number of employees, to illustrate variations in preparedness strategies across different sized power utilities. The mitigation measures listed in the figure reflect a range of proactive decisions aimed at minimizing the impact of GICs on critical grid infrastructure.

Among the most long-term adopted strategies (Figure 4 and Figure 5 subplots A), operators had "Practiced 'black-start' procedures in case of an outage" or other simulations in the highest number of cases (70%), highlighting the importance of swift power restoration in the event of system failure (although this activity is to a degree space weather agnostic). Similarly, many organizations reported having reviewed the type, condition, and location of standby spare transformers (38%), as well as utilizing past data to understand the reactive demand placed on transformers by GICs (58%). Other widely adopted measures include preparing for transformer replacements (33%), developing an inventory of vulnerable assets (21%), and ensuring that smaller structures (capacitors, converters, etc.) are adequately rated to withstand GIC effects (21%). Responses indicate a lower level of engagement with strategies such as re-adjusting negative-sequence-current protection due to higher-than-normal harmonics (0%), suggesting that some operators may prioritize physical asset readiness over protection system adjustments.

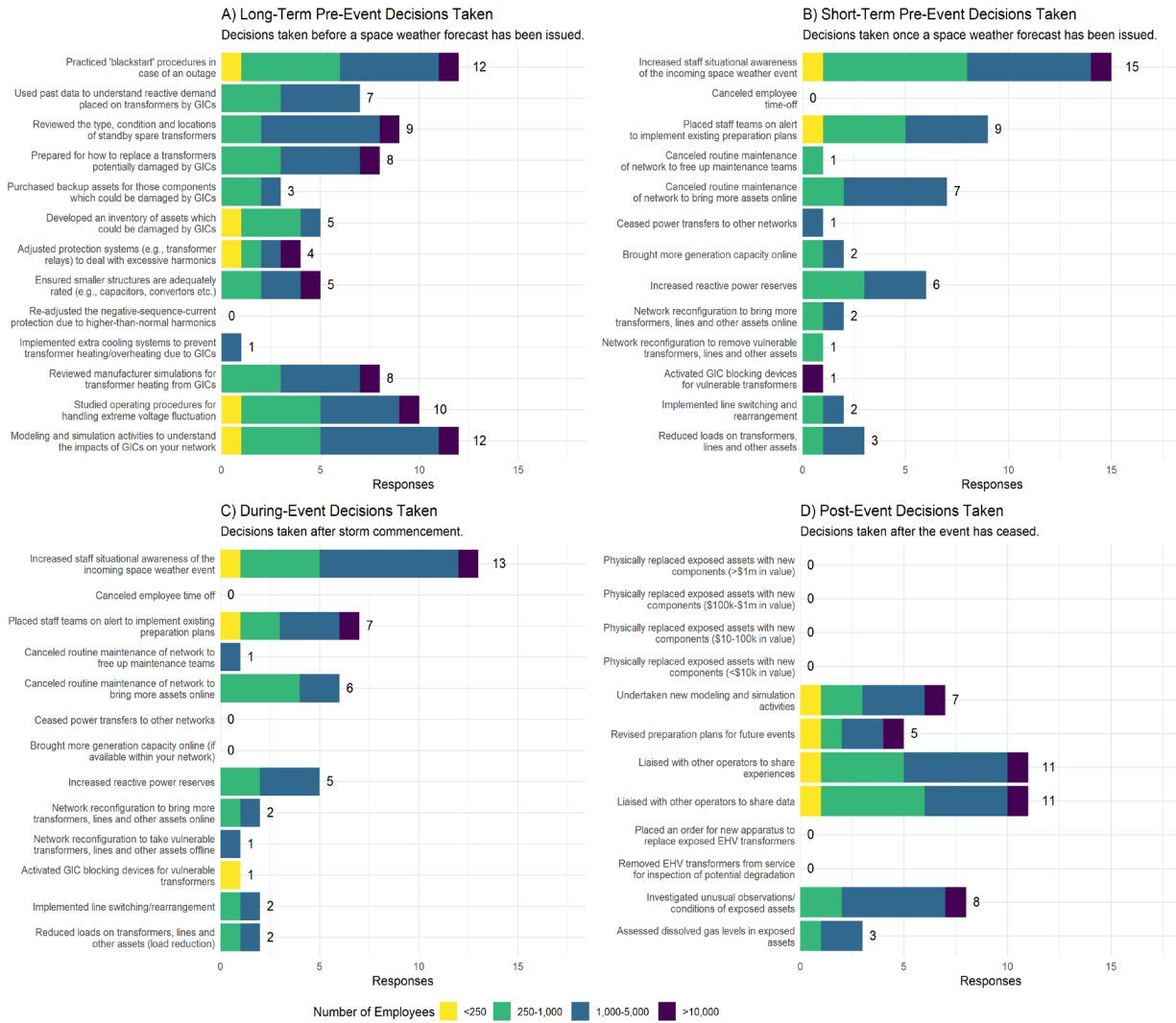

**Figure 4.** Power survey mitigation decisions taken

Notably, responses vary by organization size, with larger entities frequently engaging in modeling and simulation activities to understand GIC impacts on their networks. This pattern suggests that larger organizations may have greater resources to invest in predictive modeling and system-wide analysis. However, this may be due to larger organizations also operating higher voltage networks triggering NERC TPL-007 assessment requirements. We note that smaller organizations (<250 employees) reported fewer instances of implementing extra cooling systems or reviewing manufacturer simulations for transformer heating due to GIC exposure. This may be due to potential resource constraints in undertaking such proactive measures, or just that higher voltage assets are not in operation at smaller companies.

Next, both Figure 4 and 5 subplots' B presents survey responses from power grid operators detailing the short-term mitigation decisions, approximately 12 hours to four days before a GMD occurs, after notification from a space weather forecaster. The most reported short-term measure is "Increased staff situational awareness of the incoming space weather event" (96%), highlighting the critical role of communication and awareness in preparing for potential grid disruptions. This proactive step ensures that operators are informed and prepared to execute contingency plans as needed. A significant number of responses also indicated that utilities placed staff teams on alert to implement existing preparation plans (58%), reflecting the importance of mobilizing personnel in response to space weather forecasts. Operators commented on the existing warnings they received from NOAA ahead of solar events, one stating the "highest category of space weather warning, G5, does not provide sufficient detail about the storm severity which makes it challenging to determine necessary mitigation measures" (Id03). When staff receive notification of the warnings from NOAA, there are high levels of communication of events, but no strict action protocols based on the warnings alone. This sentiment was reflected within the results of the First National Survey of User Needs for Space Weather with summaries of sector-specific issues, including the needs for granularity of forecast products and identification of potential event impacts (Space Weather Advisory Group, 2024).

Another frequently cited action was the cancellation of routine maintenance activities (29%) or increase power reserves (25%). These strategies suggest that operators understandably prioritize maximizing system availability and workforce readiness during space weather events. However, fewer organizations reported canceling employee time-off (0%), indicating that while operational readiness is emphasized, such measures may be considered unnecessary or impractical within the short notice period.

Several grid management actions were also undertaken, including bringing additional generation capacity online (17%) and ceasing power transfers to other networks (4%). These responses indicate that while some utilities adjust power generation and flow, these measures are not as widely adopted as those focusing on personnel readiness and maintenance adjustments.

Network reconfiguration was employed to either bring more transformers, lines, and other assets online (8%) or remove vulnerable infrastructure from operation (4%). This suggests a strategic approach to balancing grid stability and asset protection, though the relatively low response counts imply that such interventions may be less commonly implemented across different organizations (e.g., those at lower latitudes).

Additional mitigation measures include the activation of GIC blocking devices for vulnerable transformers (4%), line switching and rearrangement (8%), and reducing loads on transformers and other assets (13%). These technical interventions indicate that some utilities take direct steps to limit the potential impact of GICs on critical components, though they are not as widely applied as more general preparatory actions. The responses show variation by organization size, with larger utilities more likely to report implementing technical measures such as activating GIC blocking devices or reducing loads. During an interview, one operator stressed the importance that a switching sequence had in reducing the peak current. They explained that had they not executed the sequence, the peak current would have exceeded 200A, potentially causing damage on the network (Id02). This trend suggests that larger organizations may have greater resources and flexibility to deploy protective strategies in response to space weather forecasts.

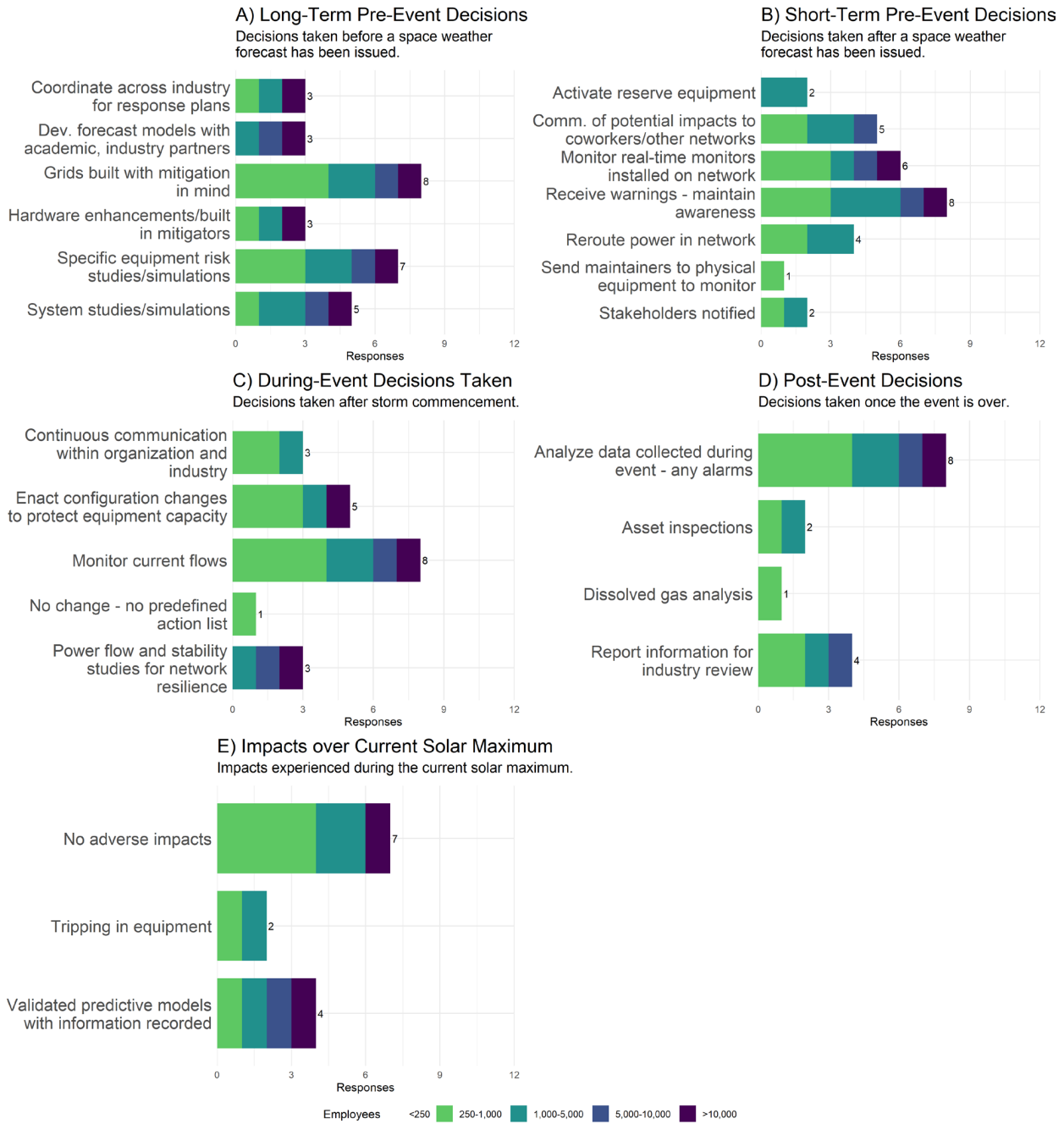

**Figure 5.** Power survey responses for mitigation decisions taken

Figure 4 and 5 subplots C present survey responses of immediate actions utilities implement to mitigate the real-time impact of GMD on grid stability. The responses are like those received for short-term pre-event decisions.

The survey findings indicate the most common reported action taken during an event was "Increased staff situational awareness of the incoming space weather event" (67%), as with short-term pre-event decisions. A significant number of responses also indicated that utilities placed staff teams on alert to implement existing preparation plans (30%), demonstrating a reliance on pre-established protocols to guide decision-making during an event. Some organizations reported canceling routine maintenance to either free up maintenance teams (4%) or bring more grid assets online (25%), indicating that ensuring infrastructure availability remains a priority even during an ongoing event. However, no organizations reported canceling employee time-off (0%) or ceasing power transfers to other networks (0%). A smaller number of operators increased reactive power reserves (21%) during the event, possibly as a means of stabilizing voltage fluctuations induced by GMDs. Additionally, a limited number of respondents engaged in network reconfiguration to bring more transformers, lines, and other assets online (37%) or take vulnerable infrastructure offline (4%). These results indicate that while grid adjustments are considered, they may be less frequently executed in real time due to operational constraints, with most decisions potentially being taken in advance.

As with short-term pre-event decisions, most grid operators primarily focus on maintaining situational awareness, mobilizing staff, and making selective adjustments to network maintenance and power reserves. The relatively low frequency of real-time grid reconfiguration or protective device activation may reflect the preference for pre-event planning and proactive mitigation rather than reactive adjustments.

Finally, for the power sector both figures' subplots D presents survey responses from power grid operators regarding the actions taken after a space weather event has concluded. These post-event decisions reflect the steps utilities undertake to assess the impact of GMDs, share insights, and improve preparedness for future occurrences.

A predominant response among operators was that they "Liaised with other operators to share experiences" (63%) and "Liaised with other operators to share data" (63%), indicating that data-sharing has played a key role in evaluating the effects of the May and October 2024 events, particularly to inform future risk management strategies.

Several operators refined their understanding of space weather impacts on grid infrastructure by stating that they had "Undertaken new modeling and simulation activities" (30%). Most operators that were interviewed agree with the survey results, stating that modeling and verification of models against recorded data helped prepare their systems for future events. Additionally, operators have "Revised preparation plans for future events" (21%) which was another commonly reported measure, demonstrating that many utilities use post-event insights to adjust their mitigation protocols and enhance resilience. Some operators "Investigated unusual observations/conditions of exposed assets" (42%), reflecting a proactive approach to identifying any unforeseen impacts. However, a smaller number of respondents reported that they had "Assessed dissolved gas levels in exposed assets" (17%), suggesting that while asset monitoring occurs, it may not be as widely implemented across all organizations.

Interestingly, no respondents indicated that they "Physically replaced exposed assets with new components" across any value category (>$1M, $100K-$1M, $10K-$100K, <$10K) (0%). This suggests that immediate asset replacement was unnecessary so far over the 2024 solar maximum.

Overall, the results emphasize that post-event actions for recent events focused on power sector collaboration via data sharing and data analysis, with the aim of refining future preparedness plans via shared understanding. The emphasis on modeling, experience-sharing, and asset inspections suggests that utilities have been focusing on learning as much as possible from these recent space weather events, with the aim of enhancing power grid resilience against future GMDs.

### 4.2. Satellite results

Figure 6 shows the long-term decisions made proactively, prior to the issuance of specific space weather forecasts, to enhance the resilience of satellite systems in both Low and Geostationary Earth Orbits.

Among the most frequently adopted strategies, the evaluation of adequate satellite shielding thickness and ongoing quality levels was reported as the most common preemptive measure (61%). 10 anonymous operators explained that many of their satellites are over-designed for solar radiation and that the radiation engineers personalize the protection for the mission and orbit, not just space weather events. Additionally, modeling and simulation approaches were widely employed, particularly for quantifying expected radiation exposure in satellite system design (55%) and for simulating orbital trajectories to assess cumulative radiation exposure (45%). One interviewee in particular commented stating that "it is all about design" when asked about their operators' long-term preparation for space weather (Id18). Other interviewees explained that not much can be done after launch, which increases the length of the design process to have five to ten years of "long-term" mitigation period (Id13, Id17).

Redundancy in satellite system design also emerged as a key mitigation strategy. Respondents indicated that they had developed redundancy design for onboard systems to safeguard against radiation exposure (48%), ensuring that critical components maintain operational functionality in adverse space weather conditions. Another strategy involved monitoring satellite asset orbital trajectories to predict which assets would experience the largest cumulative radiation exposure (26%), enabling operators to implement informed risk mitigation measures.

Ground-based infrastructure was also considered within the long-term mitigation framework. Power supply redundancy for Earth stations was implemented (45%), ensuring continued operability of control systems in the event of space weather-induced disruptions. Similarly, optical fiber link redundancy was established to enhance communication network resilience (16%), mitigating the risks of ionospheric disturbances and signal degradation.

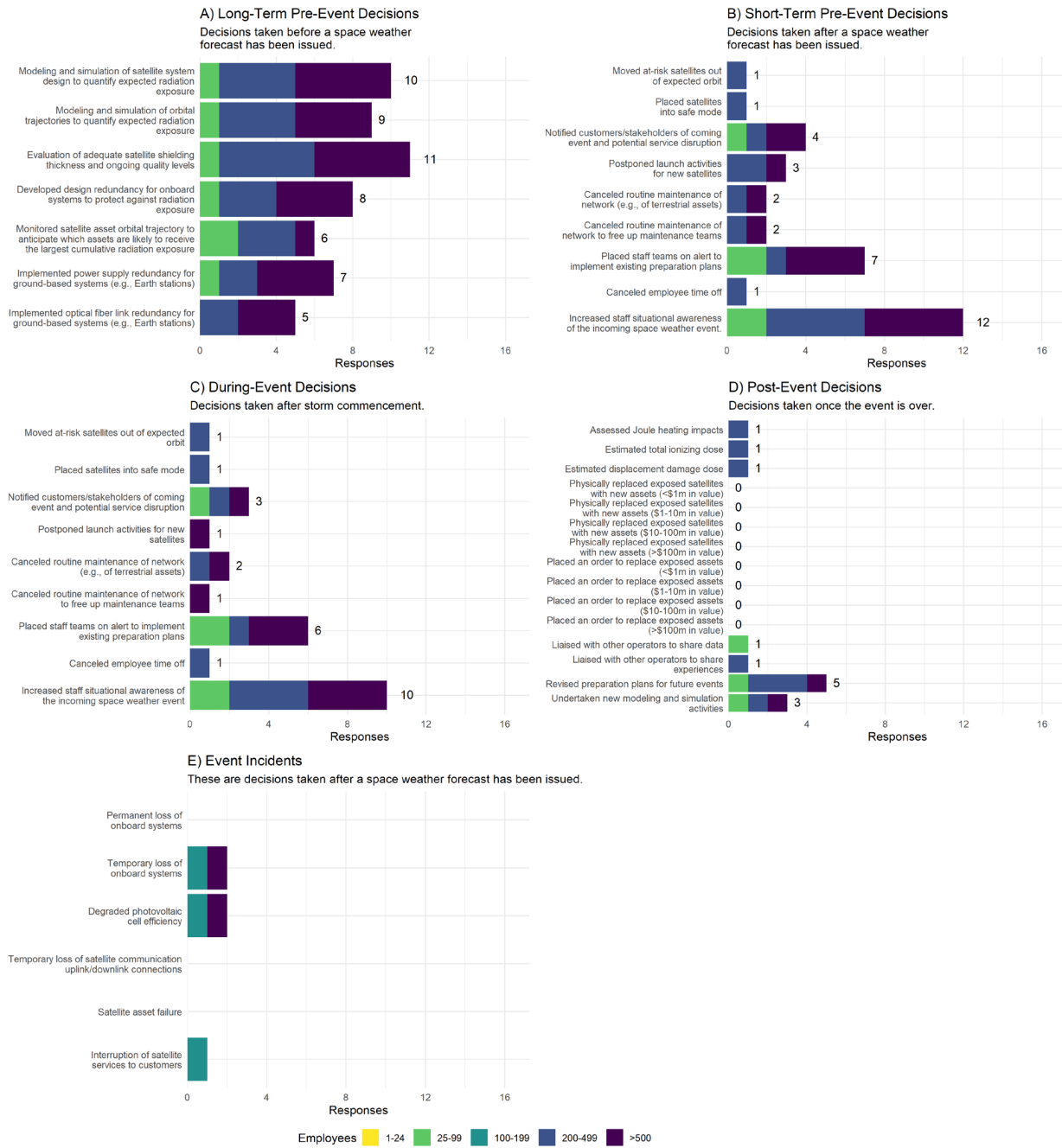

**Figure 6.** Satellite survey responses for mitigation decisions

Short-term operational decisions undertaken by satellite operators are presented in Figures 6 and 7 subplots B, in response to space weather forecasts indicating an imminent GMD. Among the most frequently reported responses, increasing staff situational awareness of the impending space weather event was the most common action (58%). This highlights the emphasis placed on ensuring personnel are prepared to monitor system performance and execute contingency plans as conditions evolve. Additionally, placing staff teams on alert to implement established preparation protocols was another widely adopted strategy (23%). During an interview, one operator explained the importance of shared understanding during events, stating that out of the history of operations during storms, with kinetic response occurring twice: repositioning the satellite to avoid reaching proton thresholds.

Communication strategies were also employed as part of the short-term response framework. Operators notified customers and stakeholders about the upcoming event and the potential for service disruptions (26%). This measure is essential for managing expectations and ensuring end-users and mission stakeholders are aware of possible performance impacts.

Several operators adjusted their operational schedules in anticipation of heightened space weather activity. Few respondents postponed launch activities for new satellites (10%), minimizing the risk of exposing newly deployed assets to elevated radiation levels or increased atmospheric drag. Additionally, operators canceled routine network maintenance to free up teams for more immediate operational tasks (6%) or suspended routine maintenance of terrestrial assets (6%), likely to reduce exposure to space weather-induced power grid fluctuations or ionospheric disturbances affecting ground communications.

Protective actions for satellite assets were less commonly reported but still noteworthy. Two operators moved at-risk satellites out of expected orbit to mitigate exposure to hazardous conditions, while another placed satellites into safe mode (10%), a precautionary step to protect onboard electronics from space weather induced anomalies. One anonymous operator explained why they avoided placing the satellites into safe mode, citing that the loss of services would outweigh the risk of damage, and that the customers almost always opted for continued service (20%).

Figures 6 and 7 subplots C presents the immediate operational responses undertaken by satellite operators once a space weather event actively impacts Earth's near-space environment. These actions reflect real-time adaptations to mitigate disruptions and protect critical satellite infrastructure during geomagnetic storms.

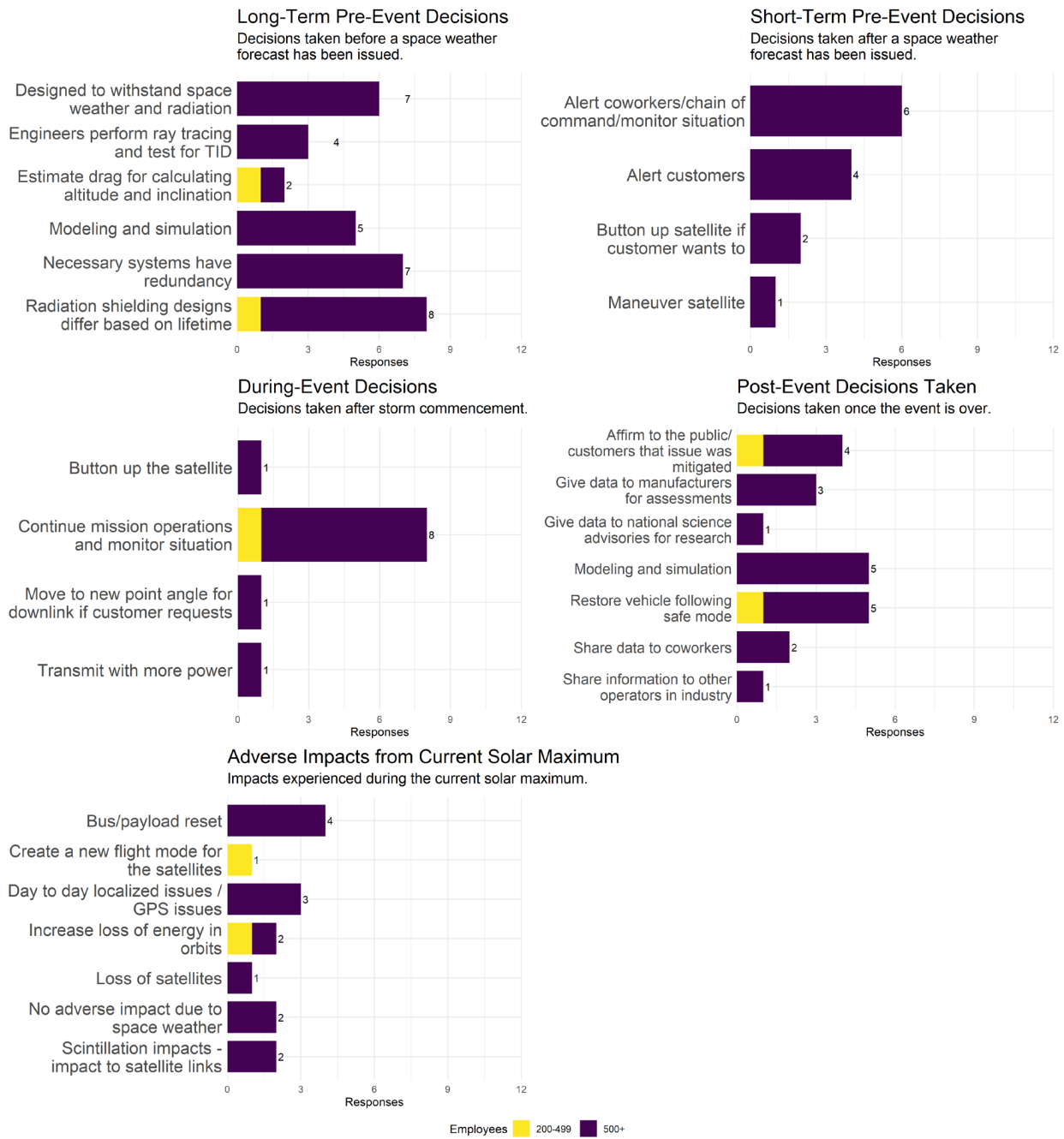

**Figure 7.** Satellite survey responses for decisions taken throughout a space weather event

The most widely reported measure involved enhancing staff situational awareness of the unfolding space weather event (32%). Additionally, operators placed their teams on heightened alert to execute pre-existing contingency protocols (19%), ensuring that personnel were prepared to intervene as necessary. Communication with external stakeholders remained an important aspect of operational decision-making, where operators notified customers and stakeholders of the ongoing event and potential disruptions to services (10%).

Operational adjustments were also made in response to the event's progression. One operator postponed satellite launch activities to avoid exposing new assets to hazardous space weather conditions (3%). Respondents reported canceling routine maintenance of terrestrial network infrastructure, a precautionary step aimed at preserving system stability and avoiding unnecessary technical interventions during a period of heightened risk (6%). Additionally, one operator freed up maintenance teams by suspending scheduled maintenance tasks, allowing personnel to focus on space weather-related contingencies (3%).

Figure subplots D illustrates the post-event decisions taken by satellite operators after a space weather event had concluded. These actions reflect efforts to assess the impacts of the disturbance, refine mitigation strategies, and enhance resilience for future solar activity. Unlike the immediate responses observed during the event, post-event decision-making is characterized by retrospective analysis, collaborative engagement, and strategic planning.

A primary focus of post-event activities was the revision of preparation plans, with operators reporting updates to their mitigation strategies based on observed system performance and response effectiveness (32%). This underscores the iterative nature of space weather resilience, where each event provides an opportunity to refine protocols and strengthen operational preparedness. Additionally, operators undertook new modeling and simulation activities, likely aimed at improving predictive capabilities and refining spacecraft vulnerability assessments (26%). In comparison, one satellite operator explained in an interview that so much time is dedicated to future planning, they often do not have the opportunity to go back and assess how accurate the previous modeling was to the events experienced during an event.

Post-event technical assessments were also conducted to evaluate the impact of the space weather disturbance on satellite assets and operational infrastructure. Operators assessed Joule heating effects, a critical factor influencing atmospheric drag and satellite orbital decay (10%), while some estimated the TID experienced by spacecraft components, which can degrade electronic performance over time (3%). A further respondent evaluated DDD, an important metric for understanding long-term material degradation due to charged particle interactions.

Collaborative efforts among operators were evident, albeit limited in scale. Two respondents liaised with other operators to share impact data (6%), while others engaged in discussions to exchange operational experiences (6%). Such collaboration plays a vital role in advancing industry-wide best practices for space weather mitigation, particularly given the interconnected nature of satellite-based services.

Notably, no operators reported taking direct asset replacement actions in response to the event. No respondents indicated having physically replaced satellites or placed orders for new spacecraft to compensate for exposure-related degradation. Some operators did explain during the interviews that they did restore the satellites in safe-mode (16%), suggesting that, at least for this solar maximum, operators did not experience immediate catastrophic failures necessitating asset replacement.

## 4.3 Aviation Sector Interview Results

Figure 8 shows all mitigation decisions made across all stages of an event for our aviation respondents. Though there were only eight interviews, the information collected is valuable to this study and recommendations.

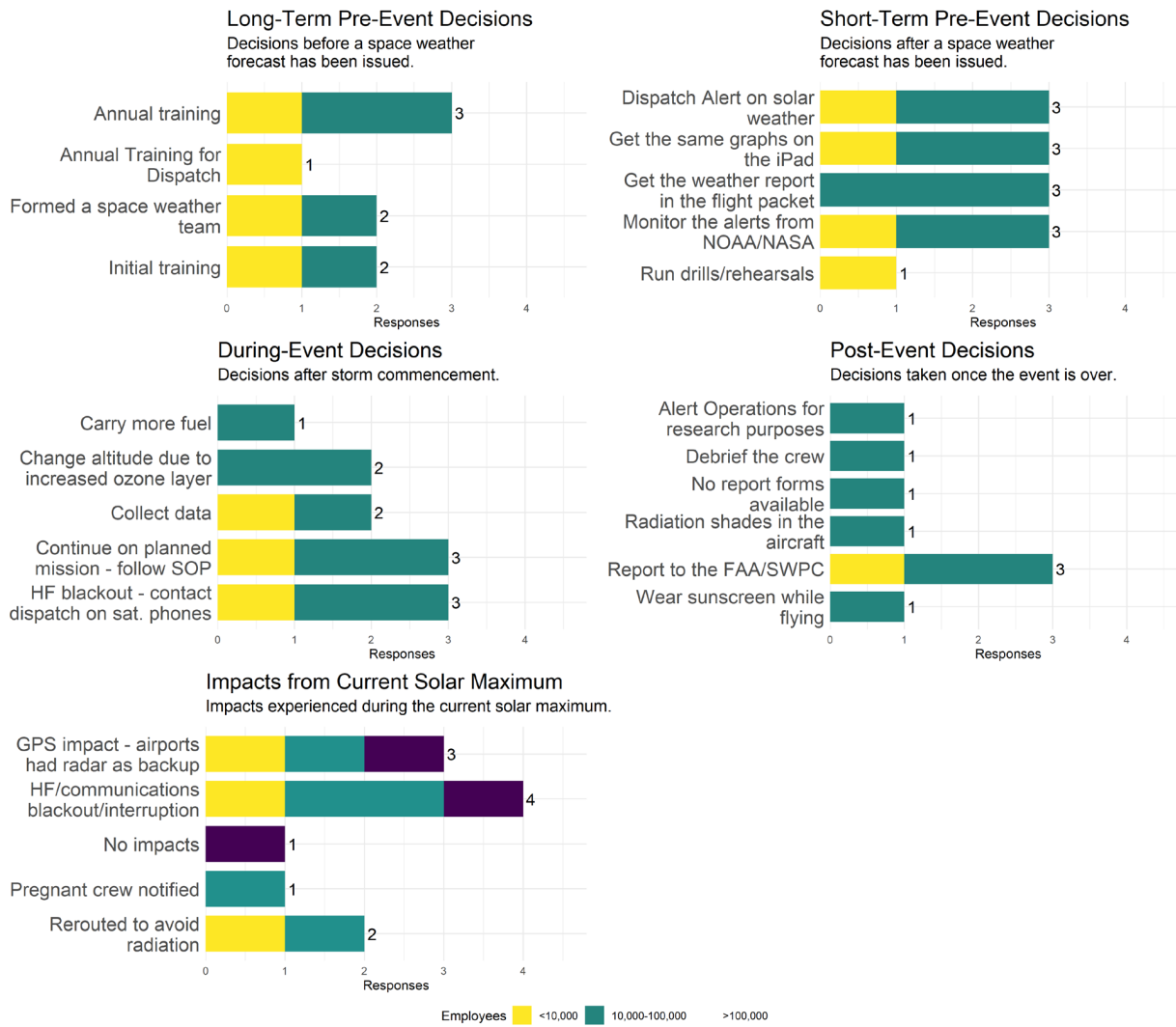

**Figure 8.** Aviation survey responses for all stages of an event

Long-term pre-event decisions for aviation were mostly focused on training, with annual (38%) and initial (25%) training both having portions focused on weather, a smaller portion on space weather impacts. One interviewee specifically mentioned that dispatch also has an annual training that includes space weather. Formations of space weather teams was also reported (25%). This phase, parallelling both power and satellite sectors, placed an emphasis on training.

Short-term pre-event decisions, shown at subplot B of Figure 8, had consistent results pointing towards shared understanding of the alerts published. Pilots reported that they would receive alerts within their flight packets (38%), alerted by dispatch during pre-flight briefings (38%), and then continue to have access to the information on their flight iPads (38%). Only one respondent reported that they would run crew rehearsals/drills on notification of an alert.

Interviewees were focused primarily on the continuation of mission during the event phase of a storm (subplot C), with no report of delays as recommended by literature. In fact, most responded to continuing mission with no change (38%), that once HF communications were impacted, then switched to satellite radios to contact dispatch (38%), and to collect data during the event (25%). Interestingly, two pilots reported changes in altitude due to an enlarged ozone layer (25%), aligning with literature recommendations. One respondent talked about carrying more fuel in response to the alerts and events (13%). This underscores the importance of retaining flexibility in actions to hazards and open communication lines to complete flights as planned.

Post-event decisions (subplot D) were dominated by one-off statements concerning alerting others to what happened and debriefing crews. The only statement to have consensus was to report the event and impacts back to the FAA and SWPC (45%), with one mention of not having the ability to report but wanting to do so. Two pilots reported changes in habit and aircraft, with inclusions of radiation shielding into aircraft (25%) and one report of increased use of sunscreen (13%).

Overall, interviewees reported instances of HF communication interruptions (50%), GPS impacts (45%), and changes to routing of flights (25%), suggesting that space weather has a play into mission planning and execution. Education is key within the aviation sector, dominating long-term pre-event and short-term pre-event phases with training and shared information. In fact, one respondent shared that they would notify any pregnant crew members or pilots of the risks of flying during an event due to increased levels of radiation.

## 5   Discussion

Large storms such as the Gannon or the October 2024 storms provide essential yet perishable data. By conducting a series of interviews and surveys, we were able to review our mitigation strategies identified through a comprehensive literature review, while documenting that essential data for future policy and network analysis.

A key portion of this study, research question one, is to identify the main mitigation strategies across three sectors in four different stages of a space weather event. All three sectors prioritized continuing mission throughout the storm, citing potential loss of capital as a driver to accept the risk of the storms. The literature pertaining to the power sector emphasizes the need for both physical assessments and changes to the network along with modeling and simulations. Most literature prescribed capital-intensive upgrades, such as expensive GIC blockers or purchasing redundancies for critical equipment (e.g., EHV transformers). Yet, this creates barriers to mitigation for smaller companies that may not have the financial ability to fund extra equipment. Certainly, we identify a range of minimal-cost mitigation measures which range from simulations to rerouting power within the grid.

The satellite industry has a unique challenge when it comes to mitigation strategies: the tyranny of distance between operators and equipment. Not surprisingly, most mitigation strategies identified were concerning designs of spacecraft, orbit, and communications. During events, it is recommended to place satellites into safe mode, maneuver out of current orbits to avoid radiation, and to increase situational awareness. Following storms, assessing damage, replacing satellites, creating new models, and sharing data within industry are the recommended actions.

The aviation industry's published mitigation strategies concern mostly pre-event and during event operations. Most recommendations focused on diversion of planned flight paths to avoid radiation exposure, with advisory systems in place to communication change in paths caused by space weather events. Pilots and crew are trained in operations and procedures for dealing with all hazards they may encounter, including solar radiation.

Research question two addressed direct impacts experienced during the solar maximum. Throughout all surveys and interviews, power, satellite, and aviation sectors stressed the importance of preparedness for space weather events. There was however a disparity between preparedness across sizes of companies, with larger companies having the resources to invest in predictive modeling, system-wide analysis, and additional resources on the networks themselves.

Surprisingly, no large purchases, repairs, or changes to networks were reported due to the Gannon storm. Most post-actions included modeling and simulations in both satellite and power sectors. The satellite surveys and interviews held emphasis on preparatory actions, since you cannot easily add equipment or change the network once in orbit. Most post-actions cited in the aviation interviews concerned updating training and notification systems for information sharing before flights. Whether it be updating other employees, stakeholders, or customers, notification about current events and actions taken are clearly necessary for ensuring success for all sizes of organizations during space weather events.

Our analysis of the differences or similarities found between the published literature and actual actions concerning space weather events addresses research question three. For satellite and power, literature recommended many more purchasing options than industry conducted when surveyed. However, both industries and literature aligned with placing emphasis in modeling both pre- and post-space weather events. Both industries expressed during qualitative interviews that they had difficulties assigning blame directly onto GIC experienced, stating that natural continuous buildup of current due to the environment may cause most of the issues and that the GIC event may cause the final push into a hazard state.

The power sector expressed resource constraints to implementing mitigation measures recommended in literature, with larger companies more likely to comply. Increased situational awareness was the most popular reported mitigation strategy conducted, as both short term pre-event and during event strategies. The industry and literature aligned on actions during events, primarily with increasing "reactive reserves and decrease loading on susceptible equipment" and increasing further situational awareness for employees and stakeholders.

When it comes to the satellite industry, both literature and practices aligned with long-term mitigation techniques. Techniques were focused on the design of the bus and the modeling of orbits and launch trajectories, with satellites designed for years of use with ample shielding and redundancy in on-board systems, whereas short-lifespan constellation satellites were designed with cost in mind. Literature recommended moving orbits or placing satellites into safe-mode, options that most operators did not take. Some operators discussed that once on mission, operators manage the system as a whole and not individual maneuvers. The literature recommends a much more risk-adverse approach to weathering geomagnetic events.

The aviation sector was underrepresented in this study but still should be noted, even with the limited number of interviews conducted. The most common statements were that pilots and crew conducted training, however decisions surrounding flights during solar events were left to the pilots for each flight. Most decisions were made prior to an event, whether it be delay, reroute, or remain on mission. There is an increased awareness throughout the community, with more pilots and crew attending training, sharing information, and creating groups to raise awareness. Most notably, one correspondent commented that pregnant pilots and crew would be less likely to take the risk of flight during solar events out of awareness of higher level of radiation (Id28). The literature published by governing bodies aligns with these comments, including increasing awareness of risk of radiation.

The benefit of collecting, analyzing and recording perishable data is the gained understanding of how vulnerable infrastructure assets are currently protected. Moreover, we often lack documented insight on the effectiveness of different mitigation decisions used to protect critical infrastructure, leaving a mismatch between actual operations and perceptions at the policy level.

The current policy context surrounding space weather sees the government keen to ensure that private sector infrastructure providers are properly prepared to guarantee services and national security. For example, the most recent National Space Weather Strategy and Action Plan (NSWSAP) was defined by the Space Weather Operations, Research, and Mitigation (SWORM) Subcommittee (SWORM 2019).

Governing bodies for each industry have a vested interest in updating models, researching and developing mitigation equipment, and facilitating sharing knowledge between companies. Current literature focuses on actions that pertain to physical changes in systems and processes, whereas industry expresses the need for precise modeling options along with more specific weather warnings. A change in equipment and configuration might not be practical, but targeted dynamic predictions from NOAA would have a large impact across industries.

Current limitations to this study include available participants from the sectors, limiting our scope to a smaller portion of each industry and may not properly represent the whole. Initially, companies from 36 countries were contacted for comments, but only 8 were represented in the surveys and interviews. Organizations may hold certain information as controlled information, thus limiting the responses or willingness to participate in the study. The nature of this unclassified study restricted wider participation, particularly government assets. It is recommended that the next steps to this continued study are to identify and increase participants from each industry, along with users who leverage GNSS in their operations.

## 6   Conclusion

This study set out to identify current mitigation options for space weather events in three industries via a systematic literature review, validate the identified options in surveys and qualitative interviews, and analyze the difference between the two sources. We contacted 303 critical infrastructure operators and received 55 responses via two mediums (surveys and interviews), with a response rate of 18%. We found that critical infrastructure operators enacted 79.2% of the mitigation options we identified via the literature. Generally, the activities implemented by critical infrastructure operators were at the lower-cost end of the range of mitigation options, suggesting there is either a limited cost-benefit incentive to invest in new protections, and/or operators are accepting potential space weather risks in order to avoid drop in service for customers. Future research needs to properly evaluate the cost-benefit ratios for the range of mitigation options identified using quantitative risk frameworks. The results of this study aim to shape the understanding of mitigation techniques and inform the industries to foster further growth in hardening critical infrastructure.